# Wave (from) Polarized Light Learning (WPLL) method: high resolution spatio-temporal measurements of water surface waves in laboratory setups

Noam Ginio[1]

*Faculty of Civil and Environmental Engineering, Technion, Haifa 320003, Israel*

Michael Lindenbaum

*Faculty of Computer Science, Technion, Haifa 320003, Israel*

Barak Fishbain

*Faculty of Civil and Environmental Engineering, Technion, Haifa 320003, Israel*

Dan Liberzon

*Faculty of Civil and Environmental Engineering, Technion, Haifa 320003, Israel*

## Abstract

Effective spatio-temporal measurements of water surface elevation (water waves) in laboratory experiments are essential for scientific and engineering research. Existing techniques are often cumbersome, computationally heavy and generally suffer from limited wavenumber/frequency response. To address this challenge, we propose the Wave (from) Polarized Light Learning (WPLL), a learning based remote sensing method for laboratory implementation, capable of inferring surface elevation and slope maps in high resolution. The method uses the polarization properties of the light reflected from the water surface. The WPLL uses a deep neural network (DNN) model that approximates the water surface slopes from the polarized light intensities. Once trained on simple monochromatic wave trains, the WPLL is capable of producing high-resolution and accurate reconstruction of the 2D water surface slopes and elevation in a variety of irregular wave fields. The method's robustness is demonstrated by showcasing its high wavenumber/frequency response, its ability to reconstruct wave fields propagating in arbitrary angles relative to the camera optical axis, and its computational efficiency. This developed methodology is an accurate and cost-effective near-real time remote sensing tool for laboratory water surface waves measurements, setting the path for upscaling to open sea application for research, monitoring, and short-time forecasting.

## 1. Introduction

Water surface waves propagate across large distances, transfer energy, induce water motion and sediment transport, and interact with shores and marine structures. Moreover, water surface waves play an important role in shaping the marine atmospheric boundary layer, affecting meso-scale and global climate processes trough air-sea mass and energy exchange (Farmer et al., 1993; Melville et al., 2002). Hence, understanding water surface waves phenomenon is crucial for environmental sciences and marine and coastal engineering. To support scientific research and applications, near-real-time monitoring and short-time projections of waves' motion and induced forces are required.

Most water waves are excited, propagate and evolve in time and space due to the wind forcing. To sample and analyze, this complex stochastic phenomenon, a spatio-temporal sensing methods are used (Harald, 2005). Specifically, one should be able to produce, high resolution, accurate measurments of the wave fields both in open sea and laboratory setups across a large range of spatio-temporal scales. One of the most important products being the wave field directional energy propagation spectrum.

---

[1] *Corresponding author*
*Email address:* noamginio@campus.technion.ac.il

Most commonly, resistance or capacitance type wave gauge probes (WGs) are used in laboratory experimental setups to retrieve reliable temporal point measurements of water surface elevation fluctuations. WGs' inherent deficiency is that they produce point measurement only. When such probes are arranged in a spatial array, one can estimate the directional energy propagation spectrum, at the location of the array (e.g. Donelan et al., 1996; Hashimoto, 1997; Toffoli et al., 2017). These methods require nontrivial interpolation, which may be both not very accurate, computationally costly and sensitive to mechanical damage and electrical interference.

Other technologies are available as well. Common acoustic techniques include pressure sensors and Acoustic Doppler Current Profilers (ADCPs) waves modules. These techniques are characterized by limited frequency response and low temporal resolution (Hashimoto, 1997). Various radar back scattering techniques are used for mapping the sea state from shores, ships, or even planes and satellites, enabling sea state assessment over large areas. However, backscattering based techniques can produce only averaged (over space or time) spectra, are expensive to implement and generally suffer from low spatial resolution (Bourdier et al., 2014; Sun et al., 2005).

Photometric techniques include ranging sensors and visible-light stereo imaging. Using these techniques for wave mapping requires site specific calibrations and costly data post processing rendering it not suitable for real time measurements (Benetazzo, 2006; Benetazzo et al., 2018, 2012; Guimarães et al., 2020; Li et al., 2021; Sun et al., 2005). Moreover, these techniques can provide only low spatial frequency version of the wave field due to the required extensive averaging (Bourdier et al., 2014; Harald, 2005).

Most of the above shortcomings are effectively addressed by a polarimetric approach, introduced by Zappa et al. (2008), utilizing a single camera armed with linear polarization filters as the sensor. This approcach is based on the physical phenomenon that unpolarized light becomes polarized when reflected from the water surface, with the reflected light polarization dependent on the local surface slope. To derive the functional relation between the local surface slope and the reflected light polarization properties Zappa et al. (2008) used a simplified physical model leading to closed-form analytical solutions of the slope. The simplified physical solution relied on two main assumptions: (1) treating the water surface as a perfect specular reflector, thus not accounting for any light up-welling from within the water body; (2) assuming overcast sky, resulting in a purely unpolarized light illuminating the water surface.

The first assumption is applicable to some coastal areas, shallow lakes, and rivers, whereas turbidity is generally very low in deep water. As for the second assumption, sun light is unpolarized at its source. However, when it reaches the Earth's surface by passing through the atmosphere, it is partialy polarized due to scattering in the atmosphere (Hannay, 2004; Smith, 2007). Therefore, under a wide range of natural light and water conditions, these simplifying assumptions as well as the analytical solution do not hold. Moreover, it is important to note that Zappa et al. (2008), demonstrated measurements of only the shortest waves in the spectrum, with wavelengths on the order of 10 centimetres.

Further studies consisting of laboratory and open sea measurments have validated the strength of the polarimetric approach but also showed it sensitivety to the illumination conditions and its inability to resolve the longer waves in the spectrum (Baxter et al., 2009; Laxague et al., 2015).

To address these issues, we propose a learning approach for deriving the polarization-to-slope (P2S) transfer function. It replaces the dependency on simplifying assumptions by the information in the learning examples. We describe the WPLL (Wave (from) Polarized Light Learning) method for producing high freuquncy/wavenumber response spatio-temporal measurements of waves fields. The method resolves the full spectrum of wavelengths. The current version of the WPLL is tailored for laboratory setups, using an artificial illumination. The WPLL method was constructed based on findings detailed in previously published feasibility study (Ginio et al., 2023),

The WPLL method includes data collection strategies and procedures including preprocessing, learning mechanisms (deep neural networks - DNN), training strategies, slope-to-surface elevation reconstruction algorithms and out-of-range (faulty) data treatment approaches.

The WPLL method is fully suited for use in laboratory wave flumes and basins. The performance of the technique is validated for a wide test set, characterized by different wave conditions, different polarimetric sensor orientations relative to the wave field propagation direction, and different spatial resolutions.

Interestingly and importantly, the proposed learning-based method may be trained on simple monochromatic wavetrains (denoted hereafter as MW), and yet can accurately reconstruct the slopes of complex irregular waves, as validated on JONSWAP waves. Moreover, the developed measuring method demonstrate high frequency and wavenumber response, otherwise unachievable with WG arrays commonly used in large wave tanks.

## 2. WPLL methodology

### 2.1 Polarimetric properties of reflected light

Water surface waves are specified by two-dimensional surface elevation fluctuatuion function $\eta(x,y,t)$, with $\nabla\eta(x,y,t)$ being the water surface slope. Due to wave field's complexity, it is often represented by its statistical properties. Commonly, a directional waves energy (density) propagation spectrum is used, obtained by the orthogonal decomposition of the wave field into spectral components. These correspond to either $f-\theta$ or $k-\theta$ depending on the available data. With $f$ being the wave frequency, $k$, being the wavenumber vector, and $\theta$ denoting the energy propagation direction (Tucker and Pitt, 2001).

The polarimetric slope sensing (PSS) concept of water waves capitalize on polarization properties of the light scattered from the air–water boundary to retrieve the instantaneous two-dimensional slope field $\nabla\eta(x,y,t)$. Zappa et al., (2008) introduced the use of a camera fitted with linear filters to measure these polarization properties, and an analytical physical model to infer the $\nabla\eta(x,y,t)$ map (Zappa et al., 2008).

The PSS geometric setting is shown in **Figure 1**. The light is reaching the image plane after being reflected from the water surface at an incident angle $\alpha$ (measured from the water surface normal). Both the incident ray, reflected ray and the water surface normal lie in the same plane denoted the plane of reflection. The incident angle $\alpha$ and the index of refraction $n$, together specify the Degree Of Linear Polarization, DOLP, of the captured light. Following Zappa et al., (2008), the DOLP can be calculated from the light intensities after linear polarization filtering as captured by the camera

(1) $\quad \text{DOLP}(\alpha, n) = \sqrt{\frac{(I_0 - I_{90})^2 + (2 \cdot I_{45} - (I_0 + I_{90}))^2}{(I_0 + I_{90})^2}}.$

Where $I_0$, $I_{45}$ and $I_{90}$ are the light intensities at 0, 45 and 90 degree linear polarization angles, as further detailed in section 2.2. As mentioned above, analytically retrieving $\alpha$ from the DOLP is achievable in ideal conditions. Note that the angle $\alpha$ is specified between the surface normal and the ray reflected to the camera, therefore the surface slope depends not only on $\alpha$ but also on the reflected ray vector.

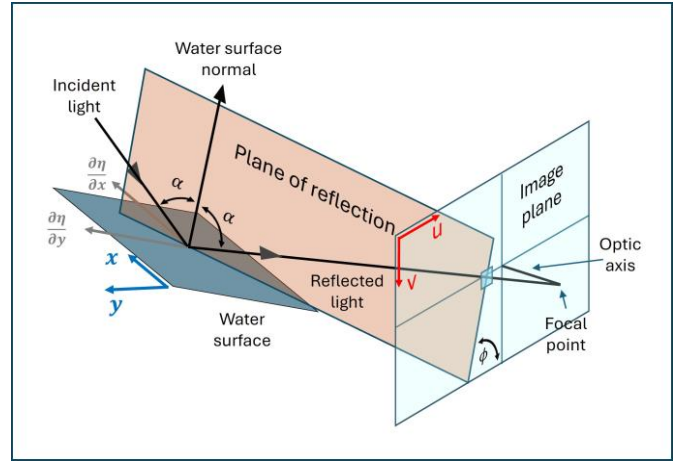

**Figure 1.** The geometric relationship between the surface normal, incidence angle relative to the water surface ($\alpha$), polarization orientation ($\phi$), and water surface slopes in x and y directions (adapted from Zappa et al. (2008)).

The inclination angle $\phi$ of the plane of reflection can be calculated from the polarization angle. Let $\Phi = 90° + \phi$. Then

(2) $\quad \Phi = \frac{1}{2} tan^{-1}\left(\frac{(2I_{45} - (I_0 + I_{90}))}{(I_0 + I_{90})}\right)$

completes the description of the surface normal orientation (Zappa et al., 2008). Note that the reflected light vector is specified by the $(u,v)$ coordinates on the image plane which correspond to $(x,y)$ coordinates at the point of reflection. The wave steepness is generally small, allowing to neglect the reflection point elevation offset from the mean water level (MWL). The projective transform between the horizontal plane at MWL and the image plane is obtained by the image geometric calibration process. Hence, $(\alpha, \phi)$ and the $(u,v)$ coordinates together specify the water surface slope, or equivalently the surface normal, at each spatial location $(x,y)$ and each time instance $(t)$.

In this work, we developed a methodology, WPLL, in which we train the DNN model that receives the light intensities filtered at four linear polarization directions in each spatio-temporal image point, as well as the spatial coordinates of that point, while providing the corresponding local water surface slopes in two dimensions.

The trained DNN model represents a weighted parametric description of the P2S transfer function. While the P2S function may depend on the light source specifications, it is invariant to the camera location and orientation relative to wave propagation direction, and to the wave field complexity.

## 2.2 Step by step calibration and measurements procedure

The overall WPLL procedure contains the following steps. Steps 1-9 are for designing and training the system and steps 10-12 are for the actual measurement:

**Calibration and training**

1. **Specify the wave field/s to be measured:**
   Characterize the wave fields to be measured and estimate the expected range of the two-dimensional water surface slopes.
2. **Set up the polarimetric camera and light source:**
   Position the polarimetric camera and the artificial light source so that the artificial light reflection as viewed by the camera covers the desired measurement region on the water surface. Once set, find the projective transform between camera's and water surface coordinate systems.
3. **Optimize light source and camera settings:**
   Configure the artificial light source power, camera's shutter speed, exposure time, and frame rate (fps) to optimize light intensity and dynamic range.
4. **Arrange and calibrate the WG probes:**
   Position the WG probes in a linear array oriented along the unidirectional *monochromatic wave train* propagation direction to be used to DNN training, covering the range (in this direction) of the intended measurement region. Perform calibration of the WG probes.
5. **Generate monochromatic wavetrains:**
   Generate unidirectional (uniform perpendicular to the wave propagation direction) MW trains of prescribed lengths and amplitudes, covering the slope values expected in the to be measured wave fields (see item **1**).
6. **Synchronize measurements:**
   Obtain records of the MW trains using the WGs array, synchronized with the camera's polarimetric imaging.
7. **Compile ground truth datasets for DNN training:**
   Deduce slopes from WG array measurements using spline derivatives. Project the slopes on the coordinate system aligned with the camera (for details see Section 3.3).
8. **Spatial averaging and data division:**
   To increase SNR, average and down-sample both light intensities and ground truth slopes on prescribed non overlapping rectangular grid *cells*. Randomly divide the *cells* into train, validation and test sets.
9. **Train the DNN:**
   Train the DNN model using backpropagation while determining the hyperparameters trough exploratory analysis using Bayesian optimization.

**Measurement**

10. **Generate the wave field to be measured:**
    Generate the wave field to be measured and collect the following data: the polarimetric camera images and synchronized *single* WG measurement (see Section 3.5).
11. **Process polarimetric inputs and utilize pretrained DNN model:**
    Down-sample the images of step 10 similarly to the process described in step 8. Feed the results into the pretrained DNN to estimate spatio-temporal slopes maps.
12. **Reconstruct surface elevation maps:**
    Use the regularized least square fit reconstruction from gradient field (grad2surf) algorithm for obtaining surface elevation maps. Correct, otherwise arbitrary, offset by a single or multiple WGs data (for details see Section 3.5).

## 2.3 Measurements setup design considerations

When designing the experimental layout for implementing this method, the first parameter to address is the range of the water surface slopes to be reconstructed. Subsequently, the spatial extent of the area of interest within the wave basin or a flume should be determined. These fundamental parameters, combined with the number of available WGs probes and the maximal region for constructing and utilizing an artificial light source above, determine the experimental setup.

The camera and the light source should be positioned in a height and angle with respect to the water surface which maximize the light reflection region from the water surface. The spatial resolution lower bound of WPLL is defined by the camera pixel resolution relative to the real-world coordinate system (RWCS), and by the selected *cell* size for spatial averaging and down sampling; see Section 3.3.4. The minimal wavenumber resolution is constrained by the area size of the light source reflection on the water surface. The temporal resolution is bounded by the camera frame rate and the frequency resolution is limited by the overall sampling time, which in turn are constrained by the computational limitation to stream, process, and store the polarimetric image data.

Once these parameters are set, supervised data for the DNN model training are collected. To maximize the ground truth accuracy and spatial extent of the supervised data region (see Section 3.1 and **Figure 2** for details), the WGs probes should be distributed evenly in an inline array along the wave basin's axis preferred for generating unidirectional MW trains. Ideally, these probes should be located adjacent to the light reflection captured by the camera.

These experimental design parameters determine the range of slopes and the final resolution and accuracy of the waves fields measured by the WPLL.

## 3. Experimental setup

In this Section we describe the experimental setup and data processing used for developing the WPLL method and testing its performance. These include instrumentation, pre and post data processing and data quality control conducted in our study.

## 3.1 Laboratory experimental instrumentation

The experiments were conducted in the Technion Sea-Air Interactions Research Laboratory (T-SAIL) wave basin. The basin dimensions were $6 \times 6\ meters$ and the water depth, $h$, was set to $70\ cm$. Flap-type mechanical wavemakers[2] (Edinburgh Designs®) were used to generate unidirectional waves. The basin was equipped with wave energy absorbing beach opposite to the wavemakers, with the sidewalls left exposed (see **Figure 2**).

Two main types of propagating waves were generated: monochromatic wave (MW) trains and irregular wave fields of a JONSWAP (Hasselmann, 1973) spectral shape, denoted as SW. The SW test datasets served as an approximation of wind-forced waves. Additionally, a set of breaking wavetrains, referred to as BW, was generated by linear focusing. The BW test dataset was employed to evaluate the system's performance on inputs corresponding to out-of-range data of the trained model (detailed in Section 3.7).

Measurements of the water surface elevation fluctuations were collected by a linear array of 8 resistance-type WGs, positioned at $12\ cm$ intervals, as shown in **Figure 2**.

Since the MW and SW generated waves are characterized by smooth water surface, we assumed continuity in both surface elevation and slope. Setting the basin coordinate $(x', y')$ as depicted in **Figure 2**, the water surface elevation $\eta(x', t)$, and slope $\nabla \eta(x', t)$ between the WGs were deduced using spline interpolation of the water surface elevation recordings at the WG locations.

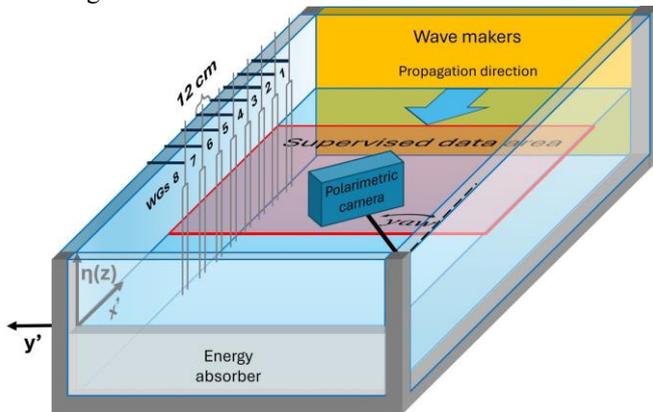

**Figure 2.** Experiment layouts of supervised data collection in the wave basin of polarized imagery depicting the spatial extent of the supervised data area bounded by the WG's locations.

The images were collected, at $32\ fps$, using a polarimetric camera (Sony IMX250MZR, $2448 \times 2048$ pixels, Polar-Mono model number BFS-U3-51S5M-C). A $12\ mm$ focal length lens was selected. Each pixel in this camera was equipped with a linear polarization filter (**Figure 3**(a)) fitted onto the camera sensor. The filters were arranged in periodic pattern, with each $2 \times 2$ pixel matrix containing filters oriented at $0°$, $45°$, $90°$ and $135°$, as shown in **Figure 3**(b, c). The four recorded light intensities, after each polarization filter referred as polarized intensities and hereafter denoted as $I_0$, $I_{45}$, $I_{90}$, and $I_{135}$, respectively.

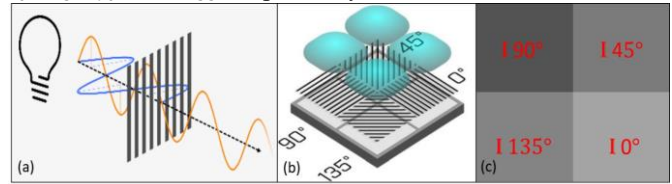

**Figure 3.** (a) The polarizing filter passes the beam that is aligned to the angle of the slits, and blocks the beam aligned perpendicular to them. (b) Each individual pixel has its own polarizing filter. (c) Zoom in image of the polarimetric image (Reproduced with permission from Teldyne FLIR).

The four reflected polarized intensities at each image point, carry the full information about the DOLP from which one can reconstruct the slope at the corresponding point on the water surface. In fact, only three polarized intensities would suffice to compute the DOLP, Eq. (1), however in this work, the slope is reconstructed directly from all polarized intensities. Note that the image point coordinates $(u, v)$ are needed for calculating the slope because they provide the required information to calculate the reflected ray (Ginio et al., 2023).

A large artificial light source, hereafter denoted as LArL, was installed above the wave basin. The use of artificial light source allowed control over the reflected light intensity and masked out ambient light reflections, thus supported high SNR. The $2 \times 2\ meter^2$ LArL source was in-house made, utilizing $24\ m$ long $12V$ flexible LED strip at $6000K$ color temperature (Inspired LED@) and was fitted with a light diffuser fabric (**Figure 4**).

---

[2] Edinburgh Designs - Wave Generator

*http://www.edesign.co.uk/product/ocean-flap-wave-generator/*

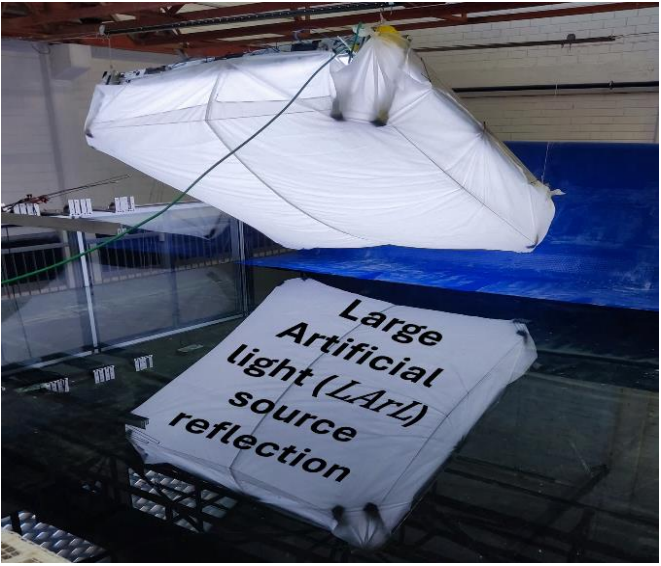

**Figure 4.** In-house made large artificial light (LArL) source reflection on the MWL. The black font notation is added in postprocessing to ease the reflection location identification by the reader.

The linear WGs array was positioned along the $x'$ axis, within the coordinates range of the LArL reflection on the water surface (**Figure 2**). WGs provided records of the instantaneous water surface elevation fluctuations at $128\ Hz$, synchronized with the camera image capturing.

The camera location (denoted $Loc$) was specified by its position and orientation relative to the wave propagation direction. Several camera locations were used in this work. One location ($Loc_A$) was used to train, validate and test the WPLL, and two additional locations ($Loc_B$ and $Loc_C$) were used for additional tests. These additional locations allowed testing the WPLL performance while measuring wave fields propagating at different angles relative to the camera. In each location a fixed image region, corresponding to the light source reflection, was cropped from the polarimetric image.

The camera locations were selected on the basin wall, at significantly different positions along the negative y' axis. At each location yaw, pitch, and elevation of the camera were set to maximize coverage of the reflected light patch at the water surface, keeping the linear WG array at the fringes of the reflection patch, while the LArL remained stationary. The camera orientation at each Loc defined new orthogonal coordinate system on the water surface plane, with the axis x being the projection of the camera optical axis, and the origin location at the y' of the corresponding Loc (See **Figure 5**).

This rotated coordinate system, is denoted hereafter as RCS. Detailed parameters are listed in **Table 1**.

**Table 1.** Camera position and orientation parameters in the three experiment repetitions ($Loc$).

| Location ($Loc$) | $Loc_A$ | $Loc_B$ | $Loc_C$ |
|---|---|---|---|
| Yaw [deg] | $-15°$ | $-30°$ | $-39°$ |
| Pitch [deg] | $-42°$ | $-35°$ | $-36.3°$ |
| Roll [deg] | $4°$ | $4°$ | $4°$ |
| x' [cm] | 0 | 0 | 0 |
| y' [cm] | -257 | -360.5 | -469.5 |
| z' [cm] | 157 | 141.4 | 160.9 |

### 3.2 Camera spatial calibration and imaging parameters tuning

To use the polarimetric data from the image coordinate system (ICS) for inferring the surface slopes in the world coordinate system (WCS) a transformation between the image plane and the MWL was needed. This was done by a projective transformation and was estimated using a standard checkerboard-based calibration process (Geiger et al., 2012). Additional projective transformations from the ICS to the various RCS-s corresponding to each camera location, $Loc_{A-C}$, were used (see Section 3.3 and **Figure 5**). The three coordinate systems ICS, WCS and RCS-s use $(u, v)$, $(x', y')$ and $(x, y)$ coordinates, respectively.

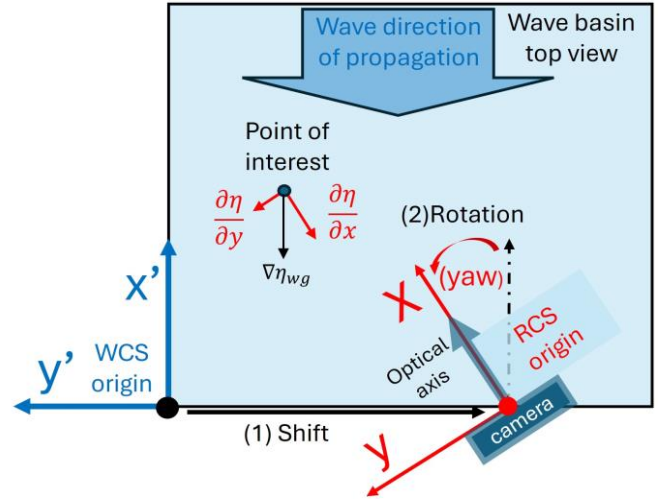

**Figure 5.** Illustration of ground truth data preperation by shift and rotation of WCS to RCS according to the camera location and orientation and vector decompostion of $\nabla \eta$.

The polarimetric camera captured the reflection of the polarized light from the water surface, produced 8-bit grayscale images. To ensure high images SNR, the controllable light source was tuned so that the camera sensor's gamma factor remained at unity and the gain was set to zero. In addition, we optimized the camera aperture and exposure time, so the dynamic range is maximized, thus, minimizing cutoff and saturation.

## 3.3 Data

### 3.3.1 Ground truth preparation

Ground truth of the measured wave field is needed not only for the MW used for training but also for MW and irregular (SW) used for testing of the method. It is important to note, that for using the WPLL as a measurement tool, a ground truth will be needed only for the monochromatic waves fields for training.

The eight WGs, positioned inline along the main propagation axis of the wavefield $x'$, recorded the surface elevation fluctuations. This data was interpolated into a continuous signal $\eta(x') = \eta(x', y'_0)$ along the $x'$ axis, using spline interpolation. The derivative of this elevation signal along x' was then computed to obtain the slope signal, $\nabla \eta(x', y'_0)$. The wave field was assumed to be perfectly unidirectional and thus uniform along y', $\eta(x', y') = \eta(x', y'_0)$, with $\frac{\partial \eta}{\partial y'} = 0$, $\nabla \eta(x', y') = \nabla \eta(x', y'_0)$. For more details, see Section 3.1. The WGs data, recorded at $128\ Hz$, were down sampled to match the image acquisition rate for each experiment by averaging sets of data points corresponding to a single frame's time span.

Next, for each camera's location, $\eta$ and $\nabla \eta$ are expressed in the corresponding RCS. That is, $\nabla \eta$ vector was decomposed into $\frac{\partial \eta}{\partial x}$ and $\frac{\partial \eta}{\partial y}$ components as shown in **Figure 5**. These two-dimensional decomposed slope vectors serve as the ground truth data for the DNN model.

Experimenting with rotated camera allows for the examination of the proposed technique (WPLL) for waves propagating at arbitrary angles relative to the ICS. These conditions are crucial for developing a model for two-dimensional slope reconstruction while generating and training the model on unidirectional wave field, with ground truth measured by a simple in-line array of point measurement probs.

### 3.3.2 Data sets selection for DNN model training

For generating the DNN training dataset, highly linear monochromatic unidirectional wavetrains were used, as they are easy to generate by the mechanical wavemaker and create a uniform wavefield in the observed area. The resolution and accuracy of the MW are primarily constrained by the distance between the WGs point measurements and the wave field linearity. Wave field linearity is crucial for constructing the ground truth elevation and slope maps within the designated supervised data region. Factors influencing wave field linearity include prescribed wave parameters, the geometry and depth of the wave basin, the mechanical wavemakers specifications, and the basin's energy absorber capabilities.

Considering our experimental setup limitation, imposed mainly by the energy absorber efficiency, long-wavelength $MWs$, with a $1\ Hz$ frequency were generated at various amplitudes to cover the range of slopes expected in the system's operational environment. This approach aims to facilitate near-real-time, one-shot reconstruction of stochastic and complex wave field test data.

In learning based functions it is important not only to train on example from the range expected in the test targets, but also to match the training example targets' distribution to that of the test. Here we train on MW targets which are characterized by a double-peaked distribution, while the irregular (SW) test slopes' pdf is of a Gaussian-like shape.

Hence, we train the DNN models on a set of local slopes in two dimensions, obtained from 12 MW trains with amplitudes ranging from $0.3\ cm$ to $3\ cm$ in $0.3\ cm$ increments. The overall distribution of this training ensemble closely matches the Gaussian-like distribution of the JONSWAP spectral shape test waves.

**Figure 6**, presents the pdfs of the slopes for three cases: the $SW_6$ experiment (see Section 3.5 for details), a single MW experiment of 3 cm amplitude, and the combined dataset of 12 MW experiments. As we can see, the SW experiment slopes pdf approximates a Gaussian shape, whereas the pdf for a single MW train exhibits a double-peaked distribution, which differ significantly. On the other hand, the slope distribution of the dataset composed of 12 MW trains, closely approximates a Gaussian-like distribution.

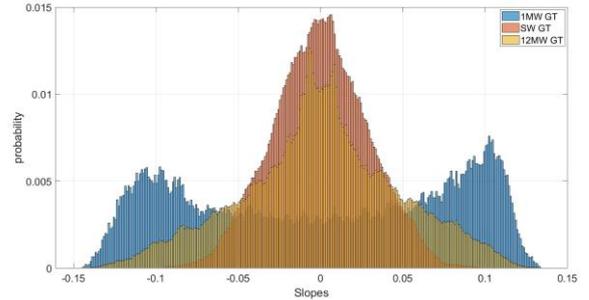

**Figure 6.** Slopes pdf of three datasets: $SW_6$ experiment, single MW experiment and 12 MW constituting the entire training set.

### 3.3.3 Irregular waves test data

The SW test sets were created by generating irregular wave fields corresponding JONSWAP energy density spectral shape, formulated as

$$(3)\quad S = \alpha_p \cdot g^2 \cdot 2\pi f^{-5} \cdot \exp\left[-\frac{5}{4} \cdot \left(\frac{f_p}{f}\right)^4\right] \cdot \gamma^{\exp\left(-\frac{(f-f_p)^2}{2\sigma^2 \cdot f_p^2}\right)}$$

where the parameters, describing the spectrum were $\omega$ denoting the angular frequency, $f = 1/T$ with $T$ being the wave period, $\gamma$ the peakiness coefficient and subscript p denoting the peak value, gravitational constant $g$, and $\sigma \in$ (0.0111 for $f < f_P$; 0.0143 for $f \geq f_P$) denoting the long and the short waves sides coefficients. $\alpha_p$ is the Phillip's equilibrium constant. The significant wave heights $H_s$ is then calculated by

$$(4) \quad H_s = 4 \cdot \left(\int_0^\infty S(f)\partial f\right)^{0.5}.$$

Specifically, the parameters chosen for our experiments are given in **Table 2**.

Whereas the training set was collected at $Loc_A$ (**Table 1**), we anticipated that P2S transfer function learned by the DNN model would not be affected by the camera orientation relative to the wave propagation direction. Thus, the same pre-trained model should be able to retrieve the two-dimensional slopes of wave fields propagating at arbitrary angle relative to the camera orientation. To test this hypothesis, the experiment were repeated with the same set of irregular waves and additional camera locations (**Table 1**).

**Table 2.** Parameters of the test set JONSWAP shaped spectrum waves (SW). the same parameters were used in each *Loc*.

| SW # | peak frequency, $f_P$ [Hz] | peak time period, $T_p$ [sec] | peakiness coefficient $\gamma$ | significant wave height, $H_s$ [cm] | dominant wavelength[3] [cm] |
|---|---|---|---|---|---|
| 1 | 0.8 | 1.25 | 10 | 2.5 | 233.66 |
| 2 | | | | 3 | |
| 3 | | | | 3.5 | |
| 4 | 1 | 1 | 15 | 2 | 155.49 |
| 5 | | | | 2.5 | |
| 6 | | | | 3 | |
| 7 | 1.5 | 0.667 | 60 | 1.5 | 69.57 |
| 8 | | | | 2 | |
| 9 | | | | 2.5 | |
| 10 | 2 | 0.5 | 100 | 1 | 39.17 |
| 11 | | | | 1.5 | |
| 12 | | | | 2 | |

### 3.3.4 Pre-processing and supervised data set assembly

The DNN model receives as an input a quadruple of pixel intensities corresponding to the four polarized intensities, $I_0$, $I_{45}$, $I_{90}$, and $I_{135}$. These light intensities are augmented by the $(u, v)$ image coordinates corresponding to the center of the $2 \times 2$ pixels matrix (**Figure 1**).

The ground truth slope components $(\frac{\partial \eta}{\partial x}, \frac{\partial \eta}{\partial y})$, also referred as targets, were synchronously coupled with each corresponding image captured by the polarimetric camera. These raw supervised datasets, intended for the DNN model training, were pre-processed as follows: Initially, all measured signals underwent visual and statistical inspection to qualitatively assess and screen out equipment malfunctions and light reflection saturation. Additionally, cross-inspection of WGs was conducted to rule out malfunctions and calibration errors. Finally, an image-WGs cross-check verified the instruments synchronization via the signal generator.

To reduce noise, square regions of pixels, denoted as *cells*, were delineated for spatial averaging. The selected *cell* size was significantly smaller than the energy carrying shortest wavelengths in the. Specifically, the *cell* size were set to non-overlapping $20 \times 20$ pixels, corresponding to about $4\ cm^2$ in WCS.

Once the *cell* extents and locations were determined, the inputs were spatially averaged over these *cells*. The corresponding cells on the water surface in the WCS were determined using the spatial calibration. For each cell in each image, the averaged input and the corresponding averaged ground truth together constituted one example. The DNN input lied in $\mathbb{R}^6$, and consisted of the four averaged polarized intensities along with the $(u, v)$ coordinates of each *cell*. The desired DNN output ground truth lied in $\mathbb{R}^2$, and contained the spatially averaged slopes $\frac{\partial \eta}{\partial x}, \frac{\partial \eta}{\partial y}$. Overall, $\#Cells \times \#Frames$ input-output pairs are available in each experiment.

### 3.4 DNN configuration and training

The selected DNN model was a multilayer feed-forward artificial neural network. The architecture and the hyperparameters of the DNN model were determined as described below (Goodfellow et al., 2016).

The network was trained on data from 12 MWs (monochromatic wavetrains, see Section 3.3.2). The training process relied on the backpropagation algorithm, which adjusted the weights and biases of the network's neurons by minimizing the root mean square error (RMSE) cost function for each prediction-target pair $i$

$$(5) \quad \text{RMSE} = \sqrt{\frac{1}{n}\sum_{i=1}^{n}\|\pi_i - \bar{\pi}_i\|^2}$$

Here $\bar{\pi}_i$ and $\pi_i$ denote the DNN model prediction and the ground truth targets of the two-dimensional water surface slopes, respectively. The summation ran on the $n$ training examples. The RMSE minimization was achieved using the Adam optimizer, a stochastic gradient descent algorithm, which employs adaptive momentum and learning rates to efficiently find the local minimum (Kingma and Ba, 2014).

To enhance generalization and prevent overfitting, two standard techniques were employed: (1) a dropout layer was added before the output layer (Srivastava et al., 2014), and (2) a standard L2 regularization term, $\lambda$, was added to the training cost function:

---
[3] According to linear dispersion relation theory

$$\min_{w \in R} RMSE + \lambda \|w\|_2, \quad (6)$$

where $\lambda$ is the regularization hyperparameter, and $w$ represents the network's learnable weights (Goodfellow et al., 2016).

The network architecture was composed of basic building blocks, each block consisting of fully connected layer, batch normalization, and hyperbolic tangent activation function. Various hyperparameters, including the number of neurons in the fully connected layer, initial learning rate, batch size, dropout rate, $\lambda$, and and the number of building blocks were examined.

To identify the optimal hyperparameters, the Bayesian optimization algorithm (Snoek et al., 2012) was applied in two stages. Initially, the algorithm ran on an initial range for each hyperparameter, determined based on previous research (Ginio et al., 2023), trial and error, and best practices from the literature (Goodfellow et al., 2016). Subsequently, the Bayesian optimization algorithm ran again on a finer range for each hyperparameter, with extended stopping criteria regarding epochs, time, and number of searches. Eventually obtaining the network optimal parameters.

The MATLAB Deep Neural Networks toolbox (Beale et al., 2018) was used to construct and train the networks. The training, validation, and test sets were all taken from $Loc_A$ MW data (**Table 1**). The three sets were constructed by randomly portioning the MW data *cell*s with 70%-15%-15% ratio. The training set was used for the backpropagation process, while the validation set was used for estimating the expected test error and regularizing the network using early stopping to prevent overfitting. Additional tests were conducted using the SW data collected at all three camera locations.

This procedure ultimately yielded the DNN model's architecture and hyperparameters that corresponded to the lowest RMSE score on the validation MW dataset. Once trained, the DNN model provided a parametric description of the approximated connections between the polarized intensities, as sensed by the polarimetric camera, and the two-dimensional slopes of the waves.

### 3.5 Post processing and performance evaluation

As mentioned above the DNN trained on MW sets, can reconstruct arbitrary, irregular wave fields. This hypothesis was tested by evaluating the model's ability to reconstruct SW test sets with various irregular wave field parameters (**Table 2**).

The final product of the WPLL is the full description of the wave field in the form of spatio-temporal maps of the water surface elevation fluctuation, $\eta(x, y, t)$. This allows examination of any needed statistical property of the wave field, such as one dimensional and directional spectra, statistical distributions of various waves parameters, etc. To complete the surface elevation reconstruction, the estimated 2D surface slopes $\nabla \eta(x, y, t)$ were integrated numerically.

$$\eta(x, y, t) = \iint \nabla \eta(x, y, t) \partial x \partial y + \eta_{wg}(x_0, y_0, t), \quad (7)$$

with the integration constant $\eta_{wg}$ obtained by a single WG point measurement of surface elevation at $(x_0, y_0)$. The constant can also be an optimal offset obtained from a surface elevation curve provided by multiple WGs array already present in the experimental setup. This constant is obtained at each time step, $t$. To estimate $\eta$ from *Eq. 7*, a regularized least-squares surface reconstruction from gradient fields algorithm (grad2surf) was implemented, solving the Sylvester equations with an additional form of regularization (Harker and O'Leary, 2015, 2013). Given the physical constraints, the raw $\nabla \eta$ outputs and the reconstructed $\eta$ were normalized to zero mean in time for each *cell* in the grid.

Once the $\eta(x, y, t)$ maps were obtained, they were processed to derive the one-dimensional frequency domain energy density spectra. In addition, the reconstructed waves' directional energy density propagation spectrum was computed. Both were then compared with spectra obtained from the WG measurements.

Given that spatial averaging and down sampling, determined by the selected *cell's* size (see Section 3.3.4), was a crucial hyperparameter within the WPLL method, additional training and evaluations were conducted to assess the model's capability and stability when the *cell* size was altered. To this end, the 12 MW training data pre-processing was repeated with spatial averaging reduced to $4 \times 4$ pixels, corresponding to approximately $0.16 \, cm^2$ in WCS.

It is important to note that the RMSE training cost function is an unnormalized score, making it unsuitable as a performance evaluation criterion across experiments with varying ground truths. Therefore, the Pearson correlation coefficients between the ground truth slopes and water surface elevation and the corresponding WPLL outputs, denoted as $R_{\nabla \eta}$ and $R_\eta$, respectively. These coefficients were used to evaluate the model's performance on the entire dataset and across the tests and all camera locations ($Loc_{A-C}$).

### 3.6 WPLL output validation

The WPLL is a spatio-temporal measurement method. As any other measurement method, the WPLL reconstruction capabilities are bounded by the experimental setup configuration. Its wavenumber resolution and wavenumber response are determined by the sample area size and the spatial sample resolution i.e., *cell* size, respectively. While the frequency resolution and frequency response are determined by the time span of the measurement and the image acquisition rate.

The frequency resolution is determined by the overall measurement time. The frequency response is limited by the temporal sampling rate. For water surface waves these limitations are easily overcome by standard equipment available in waves research laboratories.

The spatial limitations are more difficult to overcome. The wavenumber resolution is determined by the measurement region dimensions, which is bounded by the size of the artificial light source and the optical properties of the camera. The wavenumber response is determined by the spatial sampling rate. Aiming at the smallest *cell* size possible, to maintain high SNR ratio, will set the method wavenumber response limit.

When generating irregular waves test sets we observed increasing energy of higher natural harmonics with decrease in SW dominant wavelength. And appearance of standing waves in the transversal direction. This deviation from the unidirectional propagating free waves set for generation by the wavemaker are attributed to a combined effect of imperfect functionality of the energy absorber and reflection from the uncovered side walls of the basin.

These additional energy components are not properly quantified in the ground truth of the test (SW) obtained from the WG linear array measurement. First, any transversal variations are absent due to $y'$ uniformity assumption. Moreover, higher harmonics in $x'$ direction are underestimated due to finite distance between the WGs, their dimensions, and the interpolation process. These hinder validation of the WPLL outputs vs. the ground truth. To address this issue two independent evaluations were applied.

The first evaluation examined the correlation between the WPLL outputs and the ground truth as a function of spectral content. Specifically, both signals were band-pass filtered around the first three natural harmonics. Then, calculating the Pearson, $R^2$ coefficients between the (filtered) ground truth and the WPLL reconstructions of $\nabla\eta$ and $\eta$.

The second evaluation was done by imaging the water surface during selected SW tests using a standard DSLR camera. Captured images were spatially calibrated and processed to produce a light intensity mean amplitude spectrum in the wavenumber domain (Solodoch, 2024). This spectrum was then compared with the spectrum of $\eta$ produced by the WPLL method.

### 3.7 Out-of-range slope data detection and filtering out

Upon implementing the procedures described in sections 2.2 to 3.4, the WPLL is capable to produce spatio-temporal measurements of any type of waves in laboratory setups. However, as with any measurement methodology, treatment of the out-of-range input must be implemented. Here, such inputs can occur when very steep waves are formed, exhibiting slopes outside the training range, which may happen due to non-linear wave-to-wave interactions or during the appearance of breaking waves. To examine the WPLL method's response to the input corresponding to out-of-range slopes, and to develop an appropriate treatment of such data, a dedicated test case of breaking waves BW was generated (by linear focusing) and measured at $Loc_A$. A representative example of a breaking crest passing through the camera's field of view is shown in **Figure 7**.

Because the experimental design maximizes the training dataset's slope range as detailed in Section 3.1, water slopes that exceed this range often result in surface reflections of light that do not originate from the installed light source. Such reflections are captured in the images as dark patches within the illuminated supervised data area, posing two main challenges. The first challenge is the decreased SNR due to the lower light intensity, as recorded by the camera. The second challenge is the DNN inherent limitation in extrapolating outputs for conditions that fall outside the training range (Goodfellow et al., 2016). Therefore, such out-of-range data instances should be identified and flagged-out. A solution using simple binary image filter based on a light intensity threshold was proposed and is described in the Results section.

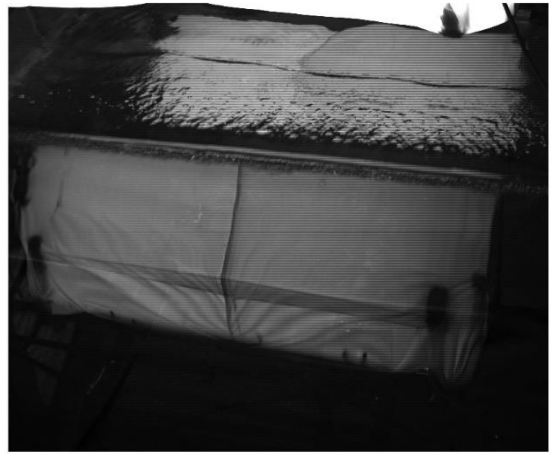

**Figure 7.** BW experiment representative example, taken at $t = 25.65\ [sec]$, depicting the capturing of light reflection out of the light source when slopes exceed the training range.

## 4. Results and discussion

### 4.1 Monochromatic propagating wave trains and DNN model training results

For the DNN model training dataset 12 MWs trains with a frequency of 1 $Hz$ and amplitudes, $a$, ranging from 0.3 $cm$ to 3 $cm$ in 0.3 $cm$ increments were generated. This resulted in a 155.47 $cm$ wavelength (see footnote 3 above), with $kh = 2.829$, and steepness $ak$ in the range 0.012 to 0.121, corresponding to a slope range of -0.1468 to 0.1353. For each wavetrain, 900 image frames were captured at 32 $Hz$ frame rate. The selected *cell*s across the 10,800 total frames yielded the MW set samples, as detailed in **Table 3**.

Table 3. Trained DNN models hyperparameters and MW dataset

| DL model name | Cell area in CCS [$pixel^2$] | Average cell area in WCS [$cm^2$] | #Cells in a frame | #Total samples | Initial learning rate | batch size | # Blocks [B] and fully connected layer size [N] | Drop-out rate | L2 regular-ization factor ($\lambda$) | # Total learnable para-meters | RMSE on MW val-idation set | $R_{\nabla\eta}$ on MW val-idation set |
|---|---|---|---|---|---|---|---|---|---|---|---|---|
| $DNN_1$ | $20 \times 20$ | 4 | 1820 | $19.656 \times 10^6$ | $4.096 \times 10^{-4}$ | $2^{13}$ | 5 B  30 N | 0.36 | $9.358 \times 10^{-6}$ | $4.292 \times 10^3$ | 0.0231 | 0.8838 |
| $DNN_2$ | $4 \times 4$ | 0.16 | 49000 | $529.2 \times 10^6$ | $7.727 \times 10^{-5}$ | $2^{15}$ | 13 B  63 N | 0.21 | $3.213 \times 10^{-7}$ | $54.749 \times 10^3$ | 0.0239 | 0.8770 |

The dataset was then split into training, validation, and test sets, following the procedure outlined in Section 3.4 and demonstrated here in **Figure 8**. The figure depicts the region of interest, superimposed on the image of the LArL refection on the water surface. Black, white and red coloured cells represent train, validation and test sets, respectively. The exploratory analysis for the preferred models hyperparameters using two different cell sizes, $DNN_1$ and $DNN_2$, was performed next, following the procedure outlined in Section 3.4. The results of this analysis are detailed in **Table 3**.

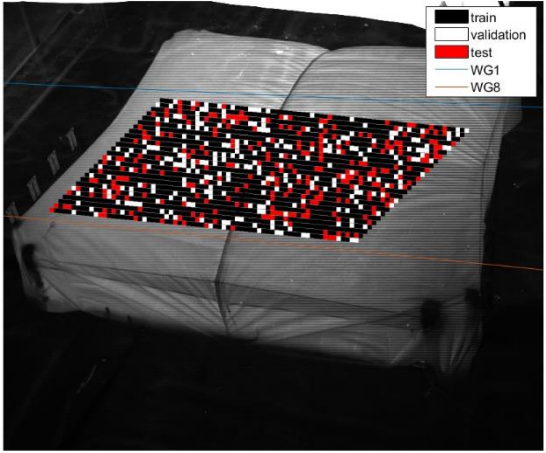

**Figure 8.** MWL image of LArL polarimetric reflection, with a *cell* selection and spatial averaging of $20 \times 20$ pixels for $DNN_1$ model training. The area between the lines is the supervised data area covered by the WGs' measurements.

### 4.2 Test sets - JONSWAP spectrum waves ($SW_{1-12}$)

To further evaluate the robustness of the WPLL reconstruction method, 12 additional JONSWAP spectra shaped waves, SW, were generated and sampled. The SW datasets were generated with the same a setup of the MW sets. A separate dataset, consists of the 12 SWs, was generated for each camera location ($Loc_{A-C}$), with camera orientation adjustment as described in Section 3.1 and detailed in **Table 1**).

These trained DNN models were used for reconstructing test sets waves, SW, without modification, testing models' capability for one-shot reconstruction of irregular and complex wave field by a network trained solely on MW dataset. The DNN models' outputs of $\nabla\eta(x,y,t)$ were then integrated numerically by solving Eq. 7 (Harker and O'Leary, 2015, 2013) to obtain the $\eta(x,y,t)$ maps for each frame. The spatial resolution achieved was approximately $2 \times 2\ cm^2$ and $0.4 \times 0.4\ cm^2$ for $DNN_1$ and $DNN_2$, respectively. A representative 3D temporal reconstruction map of $\eta$ in $SW_5$, sampled at $Loc_B$, is presented in **Figure 9**. The dots represent the ground truth surface obtained by interpolating the WG measurements, while the surface shows the $DNN_1$ model's reconstruction map of $\eta(x,y,t_0)$.

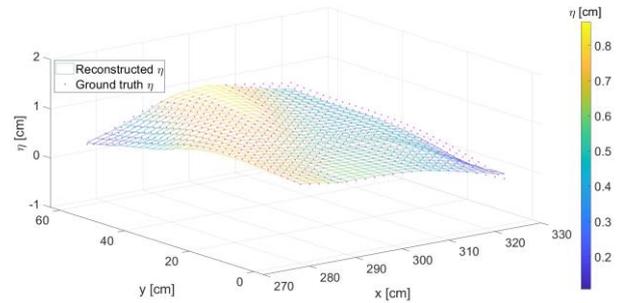

**Figure 9.** A representative example of reconstructed surface elevation map $\eta(x,y,t_0)$ integrating the $DNN_1$ provided slopes vs. the ground truth. $SW_5$ test obtained at $Loc_B$.

In our routine, the integration constant $\eta_{wg}$ for each time step was obtained by calculating the optimal surface elevation vertical correction using the WGs array data. This was achieved by aligning of the cross section of the reconstructed surface elevation with the water surface elevation curve as obtained by the WG ground truth. A representative example is shown in **Figure 10**. The black curve is the reconstructed surface elevation projected on the x'-z plane, after integrating the DNN model outputs using the grad2surf algorithm (Harker and O'Leary, 2015, 2013). Circles denote the individual WGs temporal measurements, and the green curve connecting them is the calculated spline curve. Blue arrow indicates the optimal horizontal offset correction, constituting the integration constant in eq. (7). While a single WG data point can be enough to determine the integration constant, utilizing the entire WGs array yields a more precise result. This is convenient since the array is already part of the experimental setup for generating training data.

These steps ensured accurate reconstruction of the water surface elevation, validating the DNN model's robustness and effectiveness in handling complex wave fields across different locations and orientations.

### 4.2.1 Time series reconstructions

To assess the quality of the DNN models across time, a visual inspection of water surface time series fluctuation, $\eta(x_0, y_0, t)$, was conducted. The examination was taken place at arbitrary points $\{x, y\}$ within the area of interest. The comparison was done on samples acquired from the same coordinates $[x_0, y_0] \in \{x, y\}$ on data taken from the three camera locations $Loc_{A-C}$. Representative examples of surface elevation fluctuation at the centre of the region of interest for $SW_3$ and $SW_{12}$ sets are shown in **Figure 11**(1.a) and **Figure 11**(1.b), respectively. The ground truth is marked in circles, while the integrated $DNN_1$ estimates from data sampled at $Loc_{A-C}$ are indicated by dotted curves.

In **Figure 11**(1.a), corresponding to $SW_3$, $DNN_1$ based reconstruction of the water surface elevation fluctuations of irregular waves demonstrates very high accuracy. $SW_3$ set, was characterized by a lower dominant frequency, hence a longer wavelength. On the other hand, $SW_{12}$ is of shorter dominant wavelength, as shown in **Figure 11**(1.b). In this case, the $DNN_1$ model-based reconstruction tends to overestimate both waves crests and troughs values.

These two examples illustrate the DNN model's overall tendency to effectively capture the wave period for varying wavelengths/periods, although the instantaneous accuracy may vary.

### 4.2.2 Energy density spectrum reconstruction

The quality of the spatio-temporal surface elevation fluctuations reconstructions by the WPLL was further evaluated by analyzing the wave's energy density spectrum of $\eta(x_0, y_0, t)$ time series assessed in Section 4.2.1. A

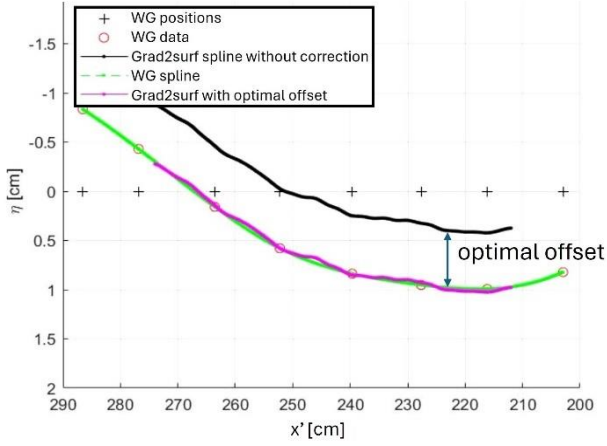

**Figure 10.** Side view on the reconstructed water surface in x'-z plane (the WG locations are marked on the MWL), depicting a single time step representative example of optimal vertical offset calculation for derivation of the reconstructed surface.

(1.a)

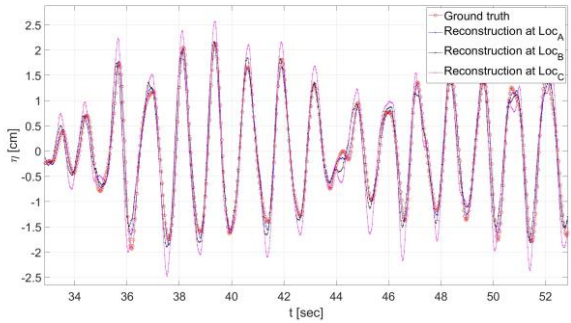

(1.b)

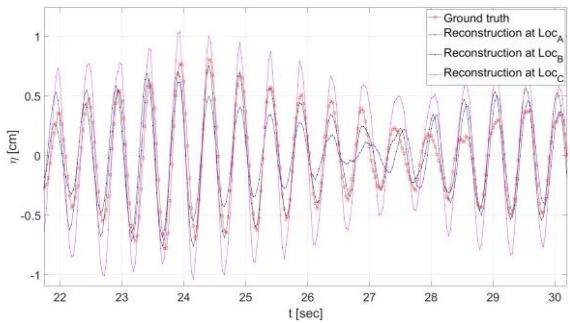

(2.a)

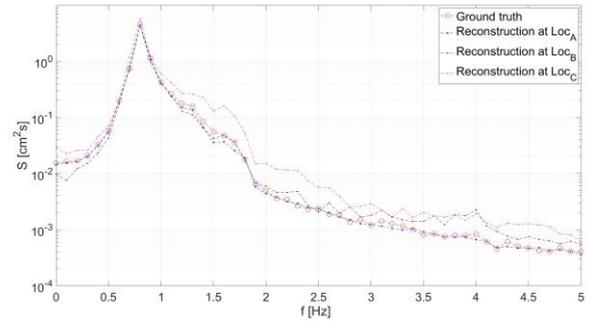

(2.b)

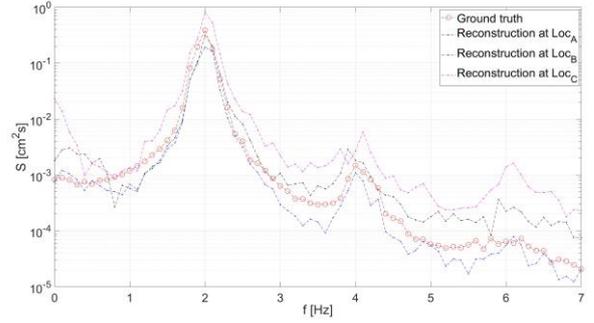

**Figure 11.** Rreconstructed instantaneous water surface elevation fluctuation, $DNN_1$, at a point located in the center of the region of interest, sampled at $Loc_{A-C}$ vs the GT. (1) Time series and (2) Energy density spectra. (a) $SW_3$ test set and (b) $SW_{12}$ test set.

comparison between the reconstructed and ground truth energy density spectra, calculated from the surface elevation fluctuations sampled at $Loc_{A-C}$, are depicted in **Figure 11**(2.a) and **Figure 11**(2.b). These representative examples are derived from the same instantaneous surface elevation fluctuation shown in Section 4.2.1.

The $DNN_1$ based reconstructed surface elevation fluctuations spectra closely match the ground truth spectra, particularly in the spectral range corresponding the free waves, carrying most of the wave's field energy around the first harmonic $f_p$ frequency. However, at higher harmonics, particularly in $SW_{12}$, presented in **Figure 11**(2.b), the $DNN_1$ based reconstructed wave spectra tend to overestimate the energy levels of the higher harmonics' peaks. This discrepancy is the result of the limited frequency and wavenumber response of the WGs array provided GT, which is thoroughly inspected and discussed in Section 4.2.5.

Consistent with the time series comparison, the spectral analysis shows that the developed technique performs better for cases with longer dominant wavelengths. Furthermore, within each $SW$ set, the accuracy of the spectral shape reconstruction, measured vs. the GT, decreases at higher frequencies.

### 4.2.3 Directional energy spectrum reconstructions

Next, the WPLL performance was examined in reconstructing the wave field spatial characteristics over the entire region of interest. To this end, directional energy density spectra were produced. A representative example of $SW_9$ (sampled at $Loc_C$) directional spectrum, averaged over the full-time span, is presented in **Figure 12**.

Obtaining data only inside the LArL reflection area, the spectral wavenumber resolution was $dk_x = 0.1208\ cm^{-1}$ and $dk_y = 0.1257\ cm^{-1}$ in the $x$ and $y$ directions, respectively. Energy peak is correctly situated at $2\ Hz$ and the wavenumber $|k|$ ranges between $0.1743\ cm^{-1}$ and $0.3487\ cm^{-1}$, within the confidence interval, estimated from the linear wave theory for a $2\ Hz$ dominant frequency and $k = 0.1606\ cm^{-1}$.

The energy propagation direction is noted to be in the range between $27.47°$ and $64.32°$ (the range bounds set by the wavenumber resolution $dk$), which again corresponds with the expected $39°$ main direction of propagation relative to the camera point of view. The wavenumber resolution is limited by our experimental setup, as it does not allow a sufficient spatial resolution. However, the energy propagation direction spread is narrow, as expected for the mechanically generated unidirectional wave field. The energy density and the directional spectra results validate the WPLL capability to reconstruct wave energy density distribution in laboratory conditions.

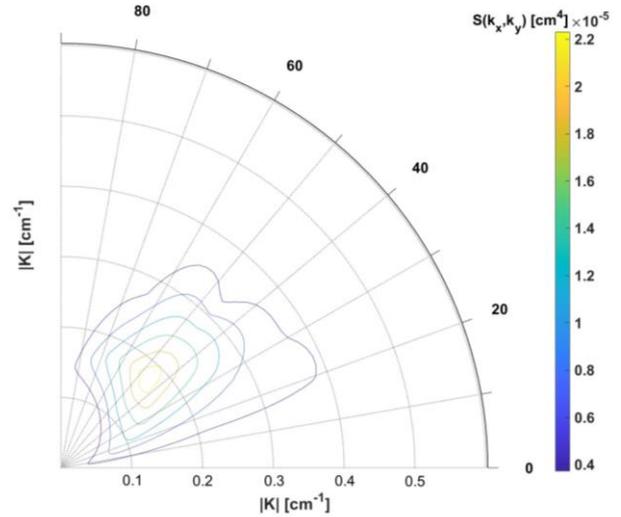

**Figure 12.** Mean directional energy density spectrum in the of reconstruction in the wavenumber domain, test set $SW_9$ sampled at $Loc_C$.

### 4.2.4 Summarizing evaluation of the WPLL method performance across $SW_\#$ and Locs

To quantitatively evaluate the new method reconstruction capabilities and compare its performance across test sets ($SW_{1-12}$) and camera locations and orientations ($Loc_{A-C}$), the metrics $R_{\nabla\eta}$ and $R_\eta$, were calculated for all SW datasets and camera locations. A total of 36 test set-$Loc$ combinations for reconstructions using the $DNN_1$ model and 36 test set-$Loc$ combinations for reconstructions using $DNN_2$ (under lower spatial averaging of $4 \times 4$ pixels). The results are listed presented in **Table 4**.

It is evident that the overall performance using the $DNN_1$ is favourable in almost every test case. When evaluating the overall performance across all tests combined, we obtain $\bar{R}_{\nabla\eta} = 0.846$ and $\bar{R}_\eta = 0.919$, with overbar denoting arithmetic mean over all test sets results. These results demonstrate the ability of the DNN model to learn well the P2S transfer function and generalize in diverse conditions. However, it is important to note that while using the $DNN_2$ model the accuracy is slightly reduced, it provides significantly higher spatial resolution using smaller *cells*. The possible reasons for the observed decrease in accuracy at higher spatial resolution are analysed and discussed in Section 4.2.5.

*Table 4. Pearson correlation coefficient, R, between the estimated $\nabla\eta$ and integrated $\eta$ outputs and the relevant ground truth for slope and surface elevation reconstructions using $DNN_1$ and $DNN_2$ models.*

|  |  | $Loc_A$ | | $Loc_B$ | | $Loc_C$ | |
| --- | --- | --- | --- | --- | --- | --- | --- |
| $SW_\#$ |  | $R_{\nabla\eta}$ | $R_\eta$ | $R_{\nabla\eta}$ | $R_\eta$ | $R_{\nabla\eta}$ | $R_\eta$ |
| 1 | $DNN_1$ | 0.836 | 0.945 | 0.816 | 0.973 | 0.816 | 0.940 |
|   | $DNN_2$ | 0.657 | 0.921 | 0.538 | 0.947 | 0.404 | 0.953 |
| 2 | $DNN_1$ | 0.861 | 0.946 | 0.832 | 0.973 | 0.823 | 0.941 |
|   | $DNN_2$ | 0.709 | 0.924 | 0.576 | 0.945 | 0.459 | 0.955 |
| 3 | $DNN_1$ | 0.876 | 0.945 | 0.843 | 0.973 | 0.828 | 0.949 |
|   | $DNN_2$ | 0.750 | 0.925 | 0.605 | 0.942 | 0.516 | 0.958 |
| 4 | $DNN_1$ | 0.858 | 0.858 | 0.829 | 0.975 | 0.819 | 0.882 |
|   | $DNN_2$ | 0.691 | 0.808 | 0.574 | 0.932 | 0.435 | 0.903 |
| 5 | $DNN_1$ | 0.878 | 0.862 | 0.837 | 0.977 | 0.836 | 0.903 |
|   | $DNN_2$ | 0.748 | 0.813 | 0.618 | 0.872 | 0.490 | 0.929 |
| 6 | $DNN_1$ | 0.896 | 0.932 | 0.859 | 0.977 | 0.822 | 0.938 |
|   | $DNN_2$ | 0.797 | 0.898 | 0.663 | 0.927 | 0.555 | 0.917 |
| 7 | $DNN_1$ | 0.890 | 0.892 | 0.856 | 0.941 | 0.855 | 0.872 |
|   | $DNN_2$ | 0.736 | 0.869 | 0.624 | 0.867 | 0.480 | 0.691 |
| 8 | $DNN_1$ | 0.911 | 0.923 | 0.849 | 0.930 | 0.847 | 0.882 |
|   | $DNN_2$ | 0.814 | 0.907 | 0.651 | 0.828 | 0.575 | 0.735 |
| 9 | $DNN_1$ | 0.906 | 0.944 | 0.869 | 0.946 | 0.843 | 0.882 |
|   | $DNN_2$ | 0.846 | 0.920 | 0.686 | 0.833 | 0.639 | 0.771 |
| 10 | $DNN_1$ | 0.879 | 0.902 | 0.835 | 0.902 | 0.822 | 0.886 |
|    | $DNN_2$ | 0.719 | 0.833 | 0.595 | 0.766 | 0.436 | 0.537 |
| 11 | $DNN_1$ | 0.845 | 0.873 | 0.798 | 0.869 | 0.806 | 0.887 |
|    | $DNN_2$ | 0.761 | 0.794 | 0.617 | 0.694 | 0.545 | 0.620 |
| 12 | $DNN_1$ | 0.880 | 0.907 | 0.808 | 0.869 | 0.803 | 0.876 |
|    | $DNN_2$ | 0.815 | 0.859 | 0.633 | 0.621 | 0.624 | 0.660 |
| Mean | $DNN_1$ | **0.876** | **0.911** | **0.836** | **0.942** | **0.827** | **0.903** |
|      | $DNN_2$ | **0.754** | **0.873** | **0.615** | **0.848** | **0.513** | **0.802** |

### 4.2.5 Re-evaluation of the ground truth data sets

As described in previous sections, we notice that WPLL reconstruction presented higher waves energy at higher harmonics, as compared to the ground truth spectra. A representative example is depicted in the energy density spectrum in **Figure 11**(2.b). At first glance such discrepancy between the reconstructed results and the ground truth data can be considered as reconstruction errors. However, the energy appears at exact multiplies of the waves field peak frequency ($f_p$), a phenomenon well expected in a wave field containing multiple harmonics due to nonlinear wave-to-wave interactions and reflection in the waves basin

As detailed in Section 3.6, this repeated quantitative observation across $SW_{1-12}$ at high frequencies, led us to examine the possibility that the observed energy at higher frequencies is indeed present in the sampled wave fields. Thus, WPLL reconstructs faithfully the wave fields, but higher harmonics are not captured by the WGs due to its limited frequency/wavenumber response. To examine this hypothesis, we have conducted several tests and additional measurements elaborated below.

First, the raw outputs of the model, $\frac{\partial \eta}{\partial x'}$ and $\frac{\partial \eta}{\partial y'}$, i.e., before integration were examined. As shown in an example case in **Figure 13**(a), the DNN model reconstructs the first and second harmonics of the $\frac{\partial \eta}{\partial x'}$ energy density spectrum closely matching those of the GT, while the third and fourth harmonics energy appears only in the DNN model reconstructed spectrum. These harmonics, of 6 Hz and 8 Hz correspond, according to the linear dispersion relation, to wavelengths of $4.9\ cm$ and $3.2\ cm$, respectively. These wavelengths are undetectable by the WGs linear array with $12\ cm$ spacing between the gauges. In addition, as per adopted methodology, the WGs array deployed along the x'-axis of the basin records zero energy in the $\frac{\partial \eta}{\partial y'}$ energy density spectrum. The DNN model outputs show small amounts of energy in the first and second harmonics of the y' slope component. This is likely due to a standing wave developing in the basin during the experiment, resulting from non-ideal energy absorption and especially the reflections from the side walls of the basin. In both cases the DNN model raw outputs are almost perfectly correspond the expected $f_p$ harmonics, clearly depicting the limitations of the WGs measurements.

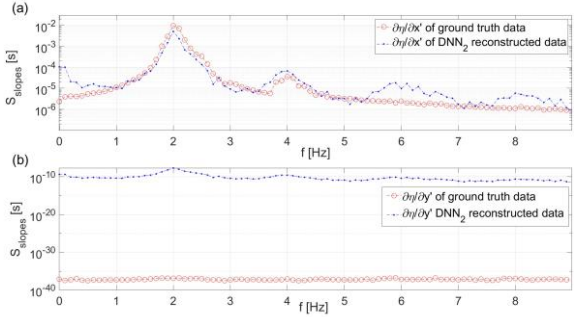

**Figure 13.** $SW_{12}$ energy density spectra, of $DNN_2$ outputs vs ground truth at $Loc_A$, of a point located in the center of the reconstructed area of (a) $\frac{\partial \eta}{\partial x'}$ and (b) $\frac{\partial \eta}{\partial y'}$.

To further assess if the differences are due to DNN model reconstruction errors or due to the limited ground truth frequency response, a per harmonic examination was performed. The $\nabla \eta$ and $\eta$ reconstructed and ground truth signals were band passed at bands corresponding to the first three harmonics of each data set. The bands limits were set to $\pm 1 Hz$ of each harmonic. This involved standard spectral decomposition using fast Fourier transform (FFT), filtering, and reverse transform on each signal. Then, for each harmonic band, the $R^2_\eta$ and $R^2_{\nabla \eta}$ were calculated. This evaluation was repeated for data sets from $Loc_{A-C}$ reconstructed using the $DNN_1$ model. A representative example of the results obtained at $Loc_A$ are presented in **Table 5**.

The $R^2$ coefficient measures the model's explained variance. For each harmonic of the sets $SW_{1-12}$, the variance explained by the model in the first harmonic is very high and is greater than the $R^2$ computed over the entire signal. For higher harmonics, the coefficient of determination decreases. The higher the harmonics the lower the coefficient. This trend is stronger in $SW_{7-12}$, which are characterized by shorter dominant wavelengths. This agrees well with the qualitative observations in **Figure 11** and visual observations of the wavefields, confirming existence of higher harmonics waves which are beyond the WGs array frequency/wavenumber response limit. While the DNN based WPLL measurements do have the high enough spectral/wavenumber response to capture shorter waves energy in all directions.

While naked eye observations did confirm this finding, observing presence of higher harmonics waves both in the main direction of waves propagation and in lateral direction, a partial quantitative confirmation was produced using common intensity images, as detailed in Section 3.6.

To investigate energy propagation and spread in the basin, an RGB, DSLR, camera, was positioned at a $40°$ degrees yaw angle to the wave propagation direction. This allowed the field of view to cover the entire wave basin surface. The LArL was removed, and the images were recorded in ambient light conditions, capturing reflections from the water surface in selected SW experiments ($SW_{3,6,9,12}$). The images of each experiment were processed by 2D FFT to produce the averaged light intensity power spectrum in the wavenumber domain.

**Table 5.** $R^2$ Pearson correlation coefficient of determination between the estimated $\nabla \eta$ and integrated $\eta$ outputs and the relevant ground truth in $DNN_1$ based reconstructions in the entire data set, and when the signal is filtered by frequency to the 1st, 2nd, and 3rd harmonics.

| $Loc_A$ | Entire signal | | 1st harmonic | | 2nd harmonic | | 3rd harmonic | |
|---|---|---|---|---|---|---|---|---|
| $SW_\#$ | $R^2_{\nabla \eta}$ | $R^2_\eta$ | $R^2_{\nabla \eta}$ | $R^2_\eta$ | $R^2_{\nabla \eta}$ | $R^2_\eta$ | $R^2_{\nabla \eta}$ | $R^2_\eta$ |
| 1 | 0.700 | 0.892 | 0.952 | 0.897 | 0.780 | 0.870 | 0.000 | 0.039 |
| 2 | 0.741 | 0.894 | 0.959 | 0.898 | 0.810 | 0.882 | 0.008 | 0.082 |
| 3 | 0.767 | 0.893 | 0.965 | 0.896 | 0.802 | 0.871 | 0.018 | 0.309 |
| 4 | 0.737 | 0.737 | 0.930 | 0.741 | 0.552 | 0.696 | 0.001 | 0.005 |
| 5 | 0.770 | 0.742 | 0.945 | 0.745 | 0.617 | 0.751 | 0.000 | 0.030 |
| 6 | 0.803 | 0.869 | 0.965 | 0.876 | 0.601 | 0.752 | 0.009 | 0.114 |
| 7 | 0.792 | 0.796 | 0.889 | 0.809 | 0.099 | 0.150 | 0.000 | 0.000 |
| 8 | 0.830 | 0.853 | 0.920 | 0.866 | 0.094 | 0.146 | 0.000 | 0.005 |
| 9 | 0.821 | 0.892 | 0.901 | 0.903 | 0.121 | 0.238 | 0.002 | 0.008 |
| 10 | 0.773 | 0.814 | 0.839 | 0.836 | 0.000 | 0.000 | 0.000 | 0.029 |
| 11 | 0.715 | 0.761 | 0.755 | 0.785 | 0.000 | 0.023 | 0.002 | 0.012 |
| 12 | 0.774 | 0.822 | 0.810 | 0.839 | 0.012 | 0.000 | 0.003 | 0.050 |

**Figure 14** shows a representative example of a raw image captured by the DSLR camera during generation of $SW_{12}$. The images are then analysed in Fourier domain. In the Fourier domain, the first frequency component is zeroed, enforcing a zero-mean signal. Which in turn eliminates energy artifacts facilitating analysis of the mean light power spectrum in the wavenumber domain derived from the water surface fluctuation.

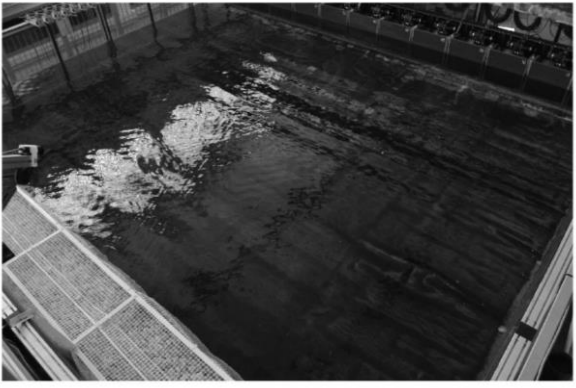

**Figure 14.** DSLR raw image covering the entire wave basin during experiment $SW_{12}$.

**Figure 15** presents the mean light intensity power spectrum in the wavenumber domain for the $SW_{12}$ DSLR images. The values are presented in logarithmic scale to increase the colormap contrast. The blue stars and black crosses represent the spectral peaks of the first four harmonics of $SW_{12}$ (**Figure 13**), shown here on the $x'$ and $y'$ axes respectively. The wavenumbers were calculated using linear dispersion relation. The red circle marks $|k| = 0.2618$ [cm$^{-1}$] from the origin, corresponding to Nyquist frequency of $12\ cm$, corresponding the WGs array spacing probes spacing. This is the WGs array wavenumber response which limit the ground truth.

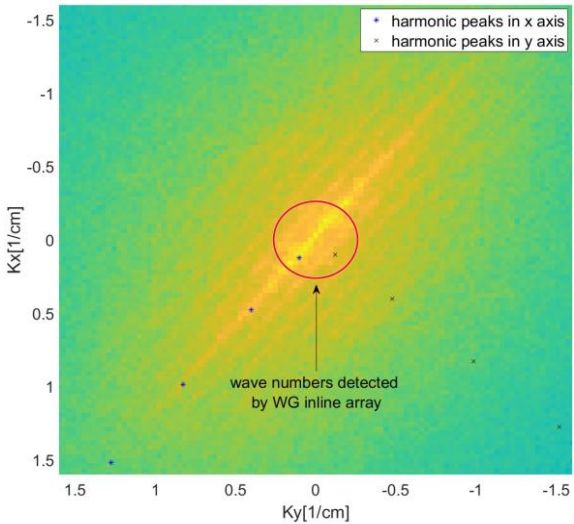

**Figure 15.** Mean light intensity power spectrum in the wavenumber domain calculated on the DSLR images of $SW_{12}$. The black crosses and blue stars represent the spectral peaks of the first four harmonics in x' and y' directions, respectively, presented in **Figure 13**. The red circle represents a distance from the origin corresponding to Nyquist frequency for spacing between WG probes.

As expected, the main wave energy propagation direction in the DSLR images' spectrum aligns with the spectral harmonic peaks on the $x'$-axis DNN model outputs (**Figure 13**(a)). This alignment corresponds to the DSLR camera orientation relative to the unidirectional wave propagation direction in the wave basin, which is 40° degrees. The $y'$-axis harmonic peaks present in the image's wavenumber spectrum, provide confirm the visual observations of the waves energy spread in this direction due to developing standing wave perpendicular to the wavetrain propagation direction.

The above analyses, supports WPLL high frequency/wavenumber response, sufficient to capture the higher harmonics energy's ability to capture high spectral/wavenumber components. However, the accuracy of higher harmonics components reconstruction cannot be fully tested quantitatively, due to the lack of sufficient information in the ground truth data. The reduction of waves energy through higher harmonics together with high accuracy of the first harmonic reconstruction suggested that the components of the waves are reconstructed with similar accuracy at all frequencies. As no alternative methodology for high accuracy sampling of spatio-temporal variations exists, one must consider that the higher harmonics are reconstructed at similar accuracy as the lower frequencies, which are shown to be very accurate, as the physical principle of light reflection polarization as a function of slope is invariant of fluctuations frequency/wavenumber and depends only on the local slope value.

### 4.2.6 DNN model performance for out-of-range slopes

The slope range of the waves one intends to measure using the WPLL must be estimated correctly before conducting the actual measurements, as described in Section 3.7. However, in some cases the measured wavefield may contain out-of-range slopes, these may occur due to the initial estimation mistakes, or due to unexpected experimental conditions generating steeper than expected waves, e.g. appearance of breakers. Hence it is crucial to understand the expected DNN model response in case it deals with polarized light input generated by an out of-range-slope, and to propose efficient mitigation strategies. For this purpose, several breaking waves experiments were staged, denoted as BW as explained in Section 3.7. The reconstruction of BW experiments water surface slopes and elevation was done using the general procedure detailed in Section 4.2. The comparison was then done at spatio-temporal location where wave breaking events were observed in the raw images (i.e. **Figure 7**). A representative example is depicted in **Figure 16**(a) and **Figure 16**(b), presenting $\eta$ and $\frac{\partial \eta}{\partial x}$ time series, respectively.

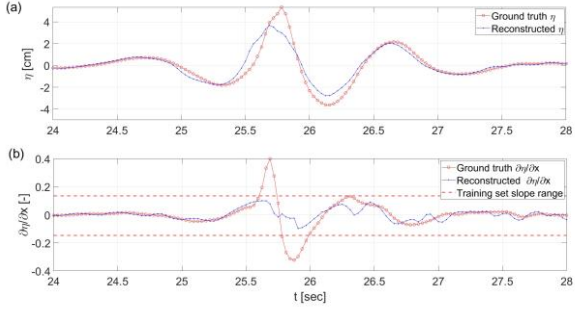

**Figure 16.** WPLL reconstructed, using $DNN_1$, time series vs the ground truth in a BW experiment at a point located in the center of the region of interest. (a) $\eta$ and (b) $\frac{\partial \eta}{\partial x}$.

The breaking crest event occurred between $25.5 - 26\ [sec]$, (presented also in the raw image in **Figure 7** taken within this time span). It is evident that the WPLL reconstruction performance deteriorates within this time range, while outside this time span, the DNN based reconstruction of both slope and elevation time series are accurate.

When examining **Figure 16**(b), during the breaking event, we can observe that the ground truth contained out-of-range slope values. The DNN underpredicts the slopes' magnitude, keeping them within the learned training range, eventually leading to wrong surface elevation reconstruction by the WPLL. It is important to note that for the highly nonlinear wave such as a breaking wave the ground truth accuracy, derived from temporal spline interpolation, is also greatly reduced.

This breaking wave scenario encompasses two main challenges to the DNN model. One challenge is the decreased SNR and dynamic range of light intensity. This happens because the breaking wave leading side is very steep and does not reflect the light emitted from the light source towards the camera. Additionally, during and after the breaking event the water surface roughness and presence of foam alter the optical properties of the water surface, introducing a significant noise to the measurements. This phenomenon is still in effect, even after the slope values return to be within the training range. This is demonstrated well in **Figure 16**(b) in time span $26 - 27\ [sec]$. The second challenge is the DNN model's inherent limitation in extrapolating outputs outside the training range data.

The combination of these challenges causes significant decrease in the system's overall reconstruction performance. Thus, effectively preventing reliable estimation of slopes and of surface elevation for out-ot-range data. Hence the strategy is to offer a robust and simple method to identify and screen out instances of out-of-range measurements.

### 4.2.7 Detection of out-of-range polarimetric Input data

To address this need, we propose using a simple and efficient, complementary tool for detecting out-of-range data, using a light intensity threshold based binary image filter. Utilizing the fact steeper slope location do not reflect the light form the light source, and hence appear as darker pixels in the images. To detect this threshold efficiently, with minimal data loss, we increased the image contrast by calculating a weighted average for each $2 \times 2$ pixels raw polarimetric image data matrix using the following formula, which was derived empirically to produce the best results for our experimental conditions:

$$(8) \quad \bar{I}_{weighted}(u,v) = (4I_0 + I_{45} + 0.25 I_{90} + I_{135})/4,$$

where, $I_0$, $I_{45}$, $I_{90}$, and $I_{135}$ are the corresponding light intensities in the $2 \times 2$ matrix at image coordinate $(u,v)$. A threshold value was then empirically selected at 140 by visually inspecting the $\bar{I}_{weighted}$ values in several representative images. Once set, this threshold served as a binary filter to detect out-of-range slopes data. **Figure 17**, presents the application of this binary filter on an image taken during a breaking wave event (shown in **Figure 7**), which was analysed in Section 4.2.6.

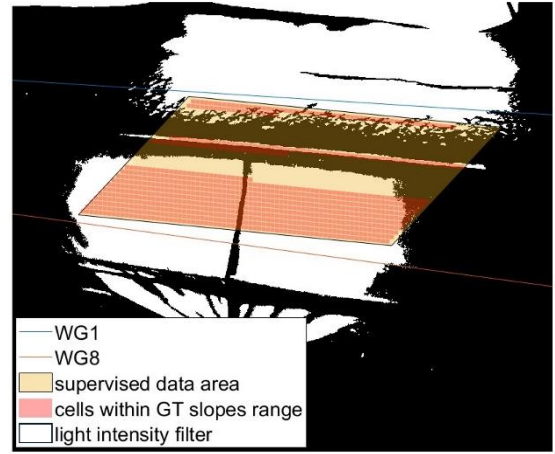

**Figure 17.** A binary threshold of 140 pixel light intensity applied on the $\bar{I}_{weighted}$ from the BW experiment image presented in **Figure 7**. The areas marked in white pixels represent *cells* identified by this threshold. The *cells* which fall within the training ground truth slope range, are highlighted in red. Additionally, the entire supervised data area, bounded by the region covered by the WGs' measurements, is delineated in yellow.

Although this is a zero approximation for the presence of out-of-range, it provides a robust and simple tool to detect and screen out data that is out-of-range, thereby increasing the performance and reliability of the system.

## 5. Conclusions

This paper introduced the Wave (from) Polarized Light Learning (WPLL) method, a remote sensing deep learning method for laboratory implementation. The WPLL was shown to be capable of inferring water surface elevation, and slope maps, in high resolution from polarimetric data of large artificial light (LArL) source reflection from the water surface.

The DNN, which is the deep learning component of the WPLL, presented high accuracy in reconstructing wave slopes from polarimetric data. The WPLL was pre-trained on polarimetric data acquired while capturing monochromatic simple wave trains (MW) from one camera location; and was capable to reconstruct complex irregular wave fields acquired from different camera locations. Thus, it presented a novel tool with strong generalization and unmatched reconstructing accuracy for a wide variety of irregular and complex waves fields.

The Pearson correlation coefficient metrics $\bar{R}_{\nabla\eta}$ and $\bar{R}_{\eta}$, presented in **Table 4**, indicated strong overall performance, with mean values of 0.846 and 0.919, respectively. These results were achieved without enforcing simplifying assumptions on incident light polarization or water turbidity, successfully resolving the entire spectrum of wavelengths. While the WPLL method effectively captures the primary energy distribution within the wave field, while also demonstrating high wavenumber and frequency response (**Figure 11**). Albeit WPLL accuracy at higher harmonics could not be assessed efficiently due to the lack of ground truth data for shorter waves components and for the transverse waves components developing in the waves basin.

Energy density spectra reconstruction aligned well with the ground truth spectra in the primary energy range. However, when compared with ground truth obtained by the WG, the WPLL method tended to overestimate energy in higher harmonics, especially in experiments with shorter wavelengths (**Figure 11**(2.b)). These discrepancies were investigated thoroughly, and we presented independent findings (**Figure 13**, **Table 5** and **Figure 15**) indicating that these deviations are not errors, but are a result of the limited wavenumber and frequency response of the wave gauges array in the main direction of waves propagation and the lack of any information in the transverse direction.

Overall, this study, successfully establishes a fully operational laboratory measurement tool, WPLL, for producing accurate spatio-temporal surface waves measurements using DNN models. This study provides future users with detailed methodology for design, implementation and quality control (QC).

Future work should focus on developing advanced filtering techniques of out-of-range inputs, to further refine the WPLL method robustness and reliability. With respect to the DNN architecture, use of convolutional neural networks, U-net and encode-decoder architectures should be investigated. Enabling incorporation of neighbouring data in the process, these architectures may improve WPLL accuracy. The presented here WPLL method resolution and frequency/wavenumber response are bounded solely by the experimental setup camera's field of view, pixel resolution and frame rate. Technological advances in computer vision and imaging, such as single-photon detectors and event cameras, indicate future leaps in frame rate acquisition and immense potential for utilization further development of this tool, paving the way for upscaling from a laboratory setup to an open sea application for research, monitoring, and short-time waves forecasting.

## Acknowledgements

The authors acknowledge the financial support provided by the Israeli Ministry of Energy (grant no.1016825) and the Israeli Ministry of Environmental Protection (grant no. 162-71). Noam Ginio is grateful for the research scholarships provided by the Israeli Port Company (IPC). A special thank goes to Ariel Weinstock and Jacob Mandel, for the aid and guidance in the construction of the large artificial light source that made this research possible.